\documentclass[12pt]{article}

\usepackage{amsmath,amsfonts,bm}

\def\eqref#1{equation~\ref{#1}}

\def\1{\bm{1}}

\DeclareMathAlphabet{\mathsfit}{\encodingdefault}{\sfdefault}{m}{sl}
\SetMathAlphabet{\mathsfit}{bold}{\encodingdefault}{\sfdefault}{bx}{n}

\newcommand{\R}{\mathbb{R}}

\DeclareMathOperator*{\argmin}{arg\,min}

\usepackage[cachedir=minted-cache]{minted}

\usepackage{caption}

\usepackage{array}
\usepackage{siunitx}

\newcommand{\mhead}[1]{%
  \rotatebox{0}{%
    \parbox{1.8cm}{\centering\scriptsize\textbf{#1}}%
  }%
}

\usepackage[utf8]{inputenc}
\usepackage{amsmath}
\usepackage{amssymb}
\usepackage{natbib}

\usepackage{ltablex}
\usepackage{xltabular}
\usepackage{booktabs}
\keepXColumns 

\usepackage{fancyhdr}
\usepackage{longtable}
\usepackage{lscape}
\usepackage{makecell}
\usepackage{threeparttable}
\usepackage[table,xcdraw, dvipsnames]{xcolor}

\usepackage[font=small,labelfont=bf,labelsep=space]{caption}
\usepackage{listings}
\lstdefinelanguage{yaml}{
  morekeywords={model,dataset,attack,evaluator,experiment_name,args,name},
  sensitive=true,
  morecomment=[l]\#,
  morestring=[b]",
  morestring=[b]',
}
\setminted{
  bgcolor=gray!10, 
  fontsize=\small  
}

\usepackage{xspace}
\usepackage[pdftex]{graphicx}
\usepackage[hidelinks]{hyperref}
\usepackage[capitalize]{cleveref}

\hypersetup{
    colorlinks = true,
    linkcolor=ACMDarkBlue,
    citecolor=ACMDarkBlue,
    urlcolor=ACMDarkBlue
}
\definecolor[named]{ACMDarkBlue}{cmyk}{1,0.58,0,0.21}

\usepackage{authblk}
\usepackage{geometry}
\usepackage{parskip} 
\usepackage[inline]{enumitem}

\usepackage{booktabs}  
\usepackage{amsfonts}
\usepackage{mathtools}
\usepackage{url}
\usepackage{enumitem}
\usepackage{comment}
\usepackage{mdframed}
\usepackage{wrapfig}
\usepackage[nottoc]{tocbibind}
\renewcommand\bibname{References}

\usepackage{ntheorem}

\definecolor{Gray1}{gray}{0.82}
\definecolor{Gray2}{gray}{0.92}

\pdfmapfile{=Cochineal.map}

\usepackage{cochineal}
\usepackage[T1]{fontenc}

\usepackage[scale=.95,type1]{cabin}
\usepackage[zerostyle=c,scaled=.94]{newtxtt}
\usepackage[cal=boondoxo]{mathalfa}
\usepackage{microtype}

\usepackage{subcaption}
\usepackage{caption}

\usepackage{fontawesome5}

\usepackage{etoolbox}

\apptocmd{\thebibliography}{\raggedright}{}{}

\usepackage{multirow}
\usepackage{tablefootnote}
\usepackage{algpseudocode}
\usepackage{algorithm}
\usepackage[pdftex]{graphicx}
\usepackage{makecell}
\usepackage{tabularx}
\usepackage{graphicx}
\usepackage{enumitem}
\usepackage{CJKutf8}
\CJKtilde

\usepackage{bera}
\usepackage{listings}
\usepackage{xcolor}
\definecolor{eclipseStrings}{RGB}{42,0.0,255}
\lstdefinelanguage{json}{
    basicstyle=\scriptsize\ttfamily,
    commentstyle=\color{eclipseStrings},
    showstringspaces=false,
    breaklines=true,
    frame=single,
    rulecolor=\color{black},
    string=[s]{"}{"},
    comment=[l]{:\ "},
    morecomment=[l]{:"},
    literate=
        *{0}{{{\color{numb}0}}}{1}
         {1}{{{\color{numb}1}}}{1}
         {2}{{{\color{numb}2}}}{1}
         {3}{{{\color{numb}3}}}{1}
         {4}{{{\color{numb}4}}}{1}
         {5}{{{\color{numb}5}}}{1}
         {6}{{{\color{numb}6}}}{1}
         {7}{{{\color{numb}7}}}{1}
         {8}{{{\color{numb}8}}}{1}
         {9}{{{\color{numb}9}}}{1}
}

\numberwithin{figure}{section}
\numberwithin{table}{section}

\usepackage{pgfplotstable} 
\usepackage{tikz}    
\usepackage{placeins}

\theoremseparator{:} 

\newmdtheoremenv[%
  backgroundcolor=white,
  linecolor=blue!60!black,
  linewidth=2pt,
  topline=true,
  rightline=false,
  skipabove=10pt,
  skipbelow=10pt,
  leftline=false]{ourexample}{Application}
  
\newmdtheoremenv[%
  backgroundcolor=gray!20,
  linecolor=red!60!black,
  linewidth=2pt,
  topline=false,
  rightline=false,
  skipabove=10pt,
  skipbelow=10pt,
  leftline=false]{ourbox}{Formulation}

\newmdtheoremenv[%
  backgroundcolor=gray!20,
  linecolor=red!60!black,
  linewidth=2pt,
  topline=false,
  rightline=false,
  skipabove=10pt,
  skipbelow=10pt,
  leftline=false]{regbox}{Box}

\theoremstyle{nonumberplain}
\newmdtheoremenv[%
  backgroundcolor=gray!20,
  linecolor=red!60!black,
  linewidth=2pt,
  topline=false,
  rightline=false,
  skipabove=10pt,
  skipbelow=10pt,
  leftline=false]{suppregbox}{Box S1}

\definecolor{quotemark}{gray}{0.7}
\makeatletter
\def\fquote{%
    \@ifnextchar[{\fquote@i}{\fquote@i[]}
           }%
\def\fquote@i[#1]{%
    \def\tempa{#1}%
    \@ifnextchar[{\fquote@ii}{\fquote@ii[]}
                 }%
\def\fquote@ii[#1]{%
    \def\tempb{#1}%
    \@ifnextchar[{\fquote@iii}{\fquote@iii[]}
                      }%
\def\fquote@iii[#1]{%
    \def\tempc{#1}%
    \vspace{1em}%
    \noindent%
    \begin{list}{}{%
         \setlength{\leftmargin}{0.1\textwidth}%
         \setlength{\rightmargin}{0.1\textwidth}%
                  }%
         \item[]%
         \begin{picture}(0,0)%
         \put(-15,-5){\makebox(0,0){\scalebox{3}{\textcolor{quotemark}{``}}}}%
         \end{picture}%
         \begingroup\itshape}%
 \def\endfquote{%
 \endgroup\par%
 \makebox[0pt][l]{%
 \hspace{0.8\textwidth}%
 \begin{picture}(0,0)(0,0)%
 \put(15,15){\makebox(0,0){%
 \scalebox{3}{\color{quotemark}''}}}%
 \end{picture}}%
 \ifx\tempa\empty%
 \else%
    \ifx\tempc\empty%
       \hfill\rule{100pt}{0.5pt}\\\mbox{}\hfill\tempa,\ \emph{\tempb}%
   \else%
       \hfill\rule{100pt}{0.5pt}\\\mbox{}\hfill\tempa,\ \emph{\tempb},\ \tempc%
   \fi\fi\par%
   \vspace{0.5em}%
 \end{list}%
 }%
 \makeatother

\usepackage{etoolbox}

\usepackage[dvipsnames]{xcolor}  
\definecolor{Blue4Head}{HTML}{004488} 
\fancypagestyle{plain}{%
  \fancyhf{}
  \fancyhead[L]{
    \raisebox{-0.15\height}{\includegraphics[height=1.2\baselineskip]{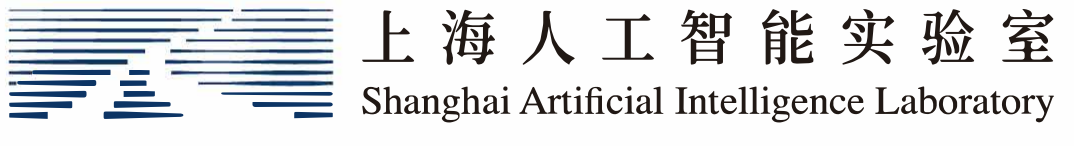}}%
  }
  \fancyhead[R]{
    \raisebox{-0.15\height}{\includegraphics[height=1.8\baselineskip]{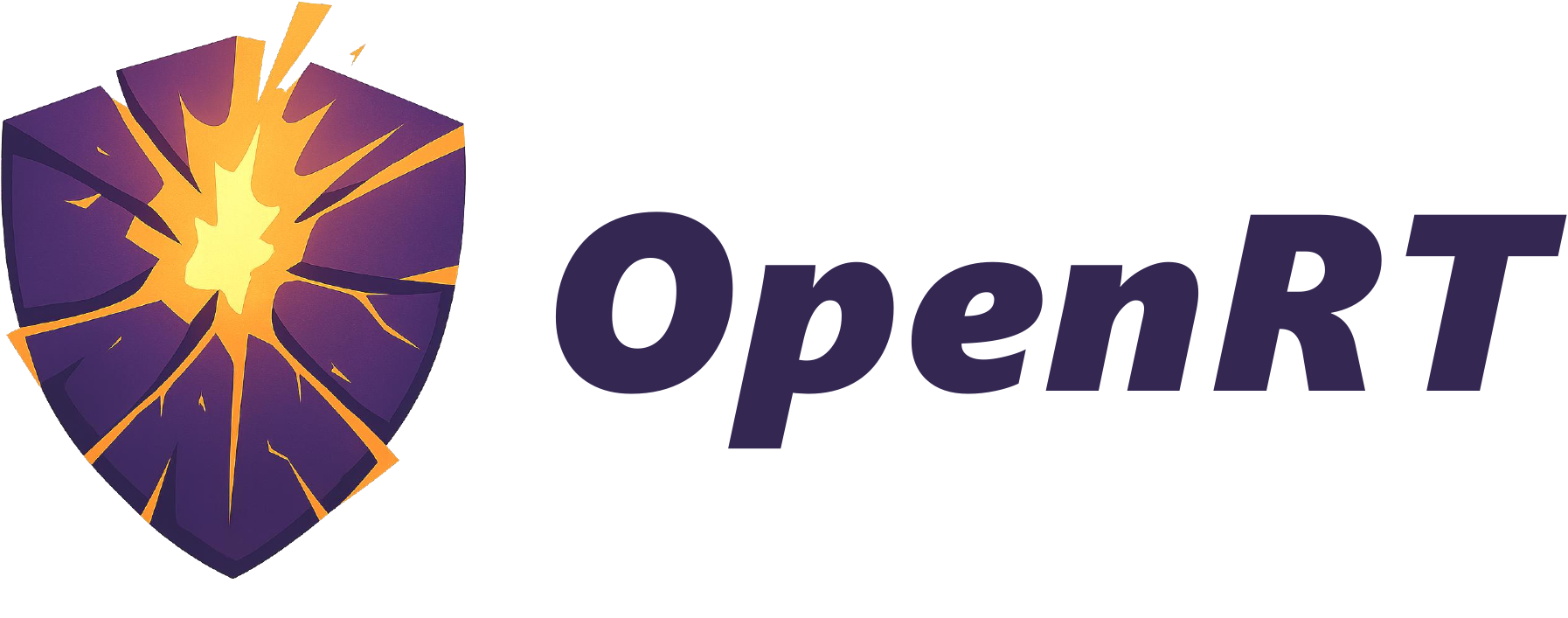}}%
  }
  
  \fancyfoot[C]{\thepage}
}

\fancyheadoffset{0pt} 
\fancyfootoffset{0pt} 

\title{\vspace{-31pt} \bf \Large OpenRT: An Open-Source Red Teaming Framework for\\Multimodal LLMs \vspace{-10pt}}

   \author[1]{Xin Wang\textsuperscript{*}}
   \author[1]{Yunhao Chen\textsuperscript{*}}
   \author[1]{Juncheng Li\textsuperscript{*}}
   \author[1]{Yixu Wang}
   \author[1]{Yang Yao}
   \author[1]{Tianle Gu}
   \author[1]{Jie Li}
   \author[1]{\authorcr Yan Teng$\textsuperscript{\textdagger}$}
   \author[1]{Yingchun Wang}
   \author[1]{Xia Hu}

\affil[1]{Shanghai Artificial Intelligence Laboratory}

\usepackage{chngcntr}
\counterwithout{equation}{section}
\counterwithout{figure}{section}
\counterwithout{table}{section}

\date{}
\begin{document}

\maketitle
  
\pagestyle{fancy}

\begin{center}
    \vspace{-26pt}
    \small
    \href{https://github.com/AI45Lab/OpenRT}{\textcolor{black!55}{\faGithub}\ \texttt{github.com/AI45Lab/OpenRT}}
    \quad$\vert$\quad
    \href{https://ai45lab.github.io/OpenRT/}{\textcolor{black!55}{\faGlobe}\ \texttt{ai45lab.github.io/OpenRT}}
\end{center}

\begin{abstract}    
    {\vspace{-8pt} \noindent
     The rapid integration of Multimodal Large Language Models (MLLMs) into critical applications is increasingly hindered by persistent safety vulnerabilities. However, existing red-teaming benchmarks are often fragmented, limited to single-turn text interactions, and lack the scalability required for systematic evaluation. To address this, we introduce OpenRT, a unified, modular, and high-throughput red-teaming framework designed for comprehensive MLLM safety evaluation. 
     At its core, OpenRT architects a paradigm shift in automated red-teaming by introducing an adversarial kernel that enables modular separation across five critical dimensions: model integration, dataset management, attack strategies, judging methods, and evaluation metrics. By standardizing attack interfaces, it decouples adversarial logic from a high-throughput asynchronous runtime, enabling systematic scaling across diverse models.
     Our framework integrates 37 diverse attack methodologies, spanning white-box gradients, multi-modal perturbations, and sophisticated multi-agent evolutionary strategies. 
     Through an extensive empirical study on 20 advanced models (including GPT-5.2, Claude 4.5, and Gemini 3 Pro), we expose critical safety gaps: even frontier models fail to generalize across attack paradigms, with leading models exhibiting average Attack Success Rates as high as 49.14\%. Notably, our findings reveal that reasoning models do not inherently possess superior robustness against complex, multi-turn jailbreaks. By open-sourcing OpenRT, we provide a sustainable, extensible, and continuously maintained infrastructure that accelerates the development and standardization of AI safety. \vspace{-7pt}
     }

\end{abstract}

\begin{center}
    \includegraphics[width=0.85\linewidth]{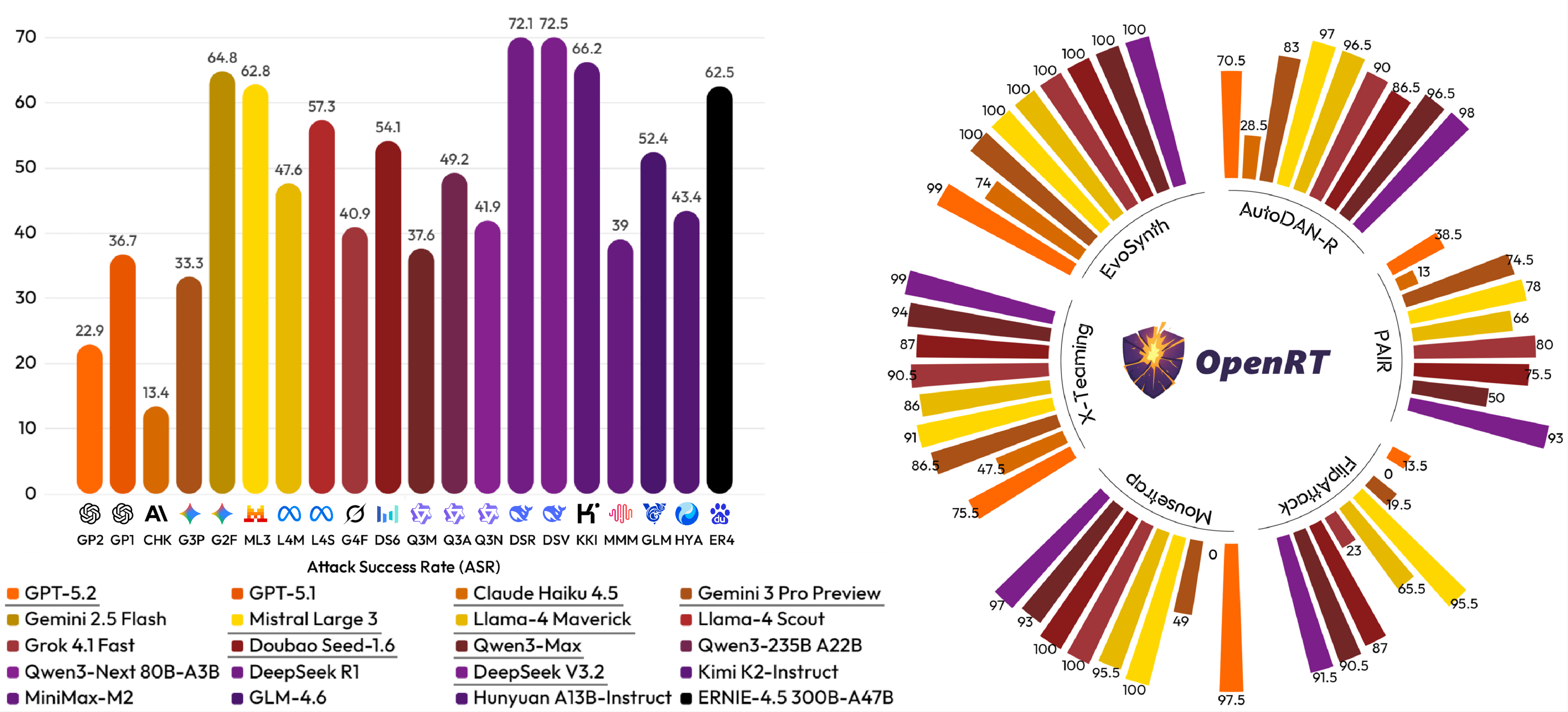}
    \captionof{figure}{\small \textbf{Left:} Average Attack Success Rate (ASR) of OpenRT across MLLMs. \textbf{Right:} Comparison against the top-6 strongest attack baselines (highest ASR) on representative MLLMs (underlined in legend).}
\end{center}

\renewcommand{\thefootnote}{}
\footnotetext{\textsuperscript{*}Equal contribution;\;
\textsuperscript{\textdagger}Correspondence regarding this technical report can be sent to \url{tengyan@pjlab.org.cn}.}

\renewcommand{\thefootnote}{\arabic{footnote}}
\setcounter{footnote}{0}

\clearpage
\pagestyle{fancy}
\section*{Takeaway Messages}

\noindent \textbf{1. Even State-of-the-Art Models Fail to Hold Ground Against Sophisticated Adversaries.} \\
Our comprehensive evaluation highlights two key findings. (1) \textbf{A clear stratification in defense capability}: Top-tier models such as Claude Haiku 4.5, GPT-5.2, and Qwen3-Max exhibit strong baseline robustness, effectively neutralizing static, template-based attacks and complex logic traps, often keeping ASR below 20\%. 
This suggests that leading labs have improved defenses against recognizable, repeatable jailbreak structures, while several models (\emph{e.g.}, Llama-4, Mistral Large 3) remain more susceptible to these simpler patterns.
(2) \textbf{A shift in the attack landscape}: adaptive, multi-turn, and multi-agent strategies dominate, whereas static, single-turn, and template-based approaches are increasingly ineffective. Methods like EvoSynth and X-Teaming can achieve $>90\%$ ASR even against advanced models. This indicates current safety training overfits to static templates, failing to generalize against the broad attack surface exposed by automated red-teaming.

\noindent \textbf{2. Adversarial Robustness Exhibits Inconsistent and Polarized Vulnerability Patterns.} \\
We observe a polarization effect where models demonstrate high resistance to specific attack families (\emph{e.g.}, text-based cipher) yet remain completely defenseless against others (\emph{e.g.}, logic nesting). For instance, Grok 4.1 Fast shows 1.5\% ASR against RedQueen but 90.5\% against X-Teaming. This stark performance disparity (${\sim}90\%$) underscores that current defenses are often patch-based rather than holistic, necessitating the multi-faceted evaluation provided by OpenRT.

\noindent \textbf{3. Enhanced Reasoning and Multimodal Capabilities are New Vectors for Exploitation.} \\
Contrary to the common assumption that more capable models are inherently safer, we find that enhanced capabilities often introduce new vectors for exploitation. Reasoning-enhanced models (CoT) do not demonstrate superior robustness; instead, their verbose reasoning processes can be manipulated to bypass safety filters. Similarly, Multimodal LLMs exhibit a critical modality gap: visual inputs frequently bypass text-based safety mechanisms, allowing cross-modal attacks to compromise models that are otherwise robust to purely textual jailbreaks. These findings suggest that current safety alignment has not kept pace with the architectural expansion of model capabilities.

\noindent \textbf{4. Proprietary Models Can Be as Vulnerable as Open-Source Models Under Certain Attacks.} \\
Our analysis reveals that proprietary and open-source models exhibit comparable susceptibility to our attack suite.
Across our 20 evaluated models, only GPT-5.2 and Claude Haiku 4.5 maintained an average ASR below 30\%, while all other models consistently exceeded this threshold. 
This universality sharply contradicts the assumption that closed deployments offer superior protection, demonstrating that the safety through obscurity of proprietary strategies fails to provide any tangible mitigation against sophisticated adversarial attacks.

\noindent \textbf{5. Scaling MLLMs Robustness via Defense-in-Depth and Continuous Red Teaming.} \\
Challenges such as polarized robustness, weak generalization to unseen attacks, and cross-modal bypasses highlight the limits of single-layer defense. 
Effective mitigation requires a paradigm shift toward Defense-in-Depth: integrating intrinsic architectural safety with runtime risk estimation and adversarial training on multimodal and multi-turn interactions. Crucially, continuous Red Teaming via infrastructure like OpenRT provides systematic evaluation to verify empirical robustness and prevent benchmark overfitting.

\clearpage

\thispagestyle{plain}
\pagenumbering{Roman}

\cleardoublepage
\setcounter{tocdepth}{3}
\tableofcontents			
\cleardoublepage

\pagenumbering{arabic}

\section{Introduction}
Multimodal Large Language Models (MLLMs)~\cite{qwen3max,gpt5-2,ClaudeHaiku45,Gemini3ProPreview,comanici2025gemini,Meta2025Llama4,Grok41Fast2025} are increasingly powering real-world applications, including conversational assistants~\cite{akhoroz2025conversational,fischer2024grillbot}, code copilots~\cite{ross2023programmer}, and search agents~\cite{gong2024cosearchagent,xi2025survey}. 
To mitigate harmful behavior, these systems are often equipped with safety alignment mechanisms~\cite{lab2025safework,qi2024safety,wang2025safevid} and safeguard policies~\cite{inan2023llama,wang2024advqdet,wang2025tapt,zhao2025qwen3guard}. Despite being widely adopted, conventional defenses such as system prompts~\cite{sharma2025sysformer}, safety filters~\cite{huang2024vaccine,gong2025safety}, and refusal-aware fine-tuning~\cite{li2024self,zhang2025safety}, remain vulnerable to adversarial attacks~\cite{chen2025evolve,rahman2025x,wang2025safeevalagent}, revealing a significant gap between perceived safety and empirical worst-case vulnerabilities.

Despite significant advances in jailbreak techniques~\cite{qi2024visual,zou2023universal,chao2023jailbreaking}, the ecosystem for systematically evaluating adversarial robustness remains fragmented. Most existing red-teaming frameworks~\cite{zhou2024easyjailbreak,chao2024jailbreakbench,mazeika2024harmbench,xu2024bag,jia2025omnisafebench} focus on a narrow subset of attacks, limited threat models, or a small selection of target models. As the number and variety of red-teaming approaches grow, including evolutionary strategies~\cite{chen2025evolve}, multi-modal jailbreaks~\cite{wang2025ideator,gong2025figstep}, multi-turn optimization~\cite{samvelyan2024rainbow,liu2024autodan}, and multi-agent coordination~\cite{rahman2025x,ren2024derail}, the lack of a unified experimental framework has become a critical bottleneck. Such fragmentation undermines reproducible benchmarking and limits the systematic evaluation of vulnerabilities across models. As a result, it remains difficult to establish standardized baselines for attack efficacy or quantify the consistency of safety failures across diverse models.

In this work, we introduce \textbf{OpenRT}, a modular and extensible framework designed for the red teaming of MLLMs. Unlike existing toolboxes often limited to a narrow subset of classic attacks, OpenRT supports massively parallel jailbreaking in both white-box and black-box settings, engineered specifically for high-throughput evaluation (Table \ref{tab:comparison}). Architecturally, OpenRT functions as a composable toolkit that explicitly decouples core components: models, datasets, attacks, judges, and evaluators under a central orchestrator. 
The framework integrates 37 attack implementations, covering a broad spectrum of threat models. In the black-box settings, OpenRT supports diverse methodologies, ranging from direct single-turn prompting~\cite{liu2023autodan,liu2024flipattack} to sophisticated multi-turn conversational jailbreaks (\emph{e.g.}, PAIR~\cite{chao2023jailbreaking}, RedQueen~\cite{jiang2024red}, Crescendo~\cite{russinovich2025great}, RACE~\cite{ying2025reasoning}), code-oriented exploitation (\emph{e.g.}, CodeAttack~\cite{ren2024codeattack}), and population-based optimization (\emph{e.g.}, genetic algorithms, GPTFuzzer~\cite{yu2023gptfuzzer}). It also leverages multi-agent and diversity-driven approaches, such as X-Teaming~\cite{rahman2025x}, Rainbow Teaming~\cite{samvelyan2024rainbow}, and EvoSynth~\cite{chen2025evolve}, to maximize the exploration of jailbreak trajectories. In the white-box settings, the framework facilitates gradient-based attacks, including GCG~\cite{zou2023universal}, visual perturbations~\cite{qi2024visual}, and imperceptible jailbreaking~\cite{gao2025imperceptible}. Powered by an asynchronous engine, it unifies API and local model interfaces for scalability. Furthermore, it employs a hybrid evaluation suite that combines rule-based filters with LLM judges, ensuring efficient and robust assessment across diverse safety domains.

We demonstrate the utility of OpenRT through a comprehensive benchmark of 20 distinct MLLMs, represented by advanced models such as GPT-5.2, Claude Haiku 4.5, Gemini 3 Pro Preview, Qwen3-Max, Doubao-Seed-1.6, and DeepSeek-V3.2. Our experiments expose critical safety vulnerabilities in current deployments, revealing a widespread average Attack Success Rate (ASR) of 49.14\%. Notably, even advanced models remain susceptible, exhibiting ASRs ranging from 13.4\% to as high as 72.5\%. These findings highlight the significant fragility of existing safeguards and signal an urgent need for more robust defense mechanisms.

In summary, our main contributions are:
\begin{itemize}
    \item We introduce \textbf{OpenRT}, a modular framework that unifies fragmented attack methods into a standardized orchestration system. By decoupling adversarial logic from a high-concurrency asynchronous runtime, OpenRT enables high-throughput parallel evaluation and streamlines the deployment of complex, multi-agent, and multi-modal attack scenarios at scale.
    
    \item We integrate \textbf{37 diverse attack algorithms} spanning both white-box and black-box threat models. These encompass multi-turn conversational strategies, multi-modal jailbreaking, and multi-agent coordination.
        
    \item We conduct an extensive empirical study across \textbf{20 advanced MLLMs}. Our results reveal that even frontier models, including GPT-5.2, Claude Haiku 4.5, Gemini 3 Pro Preview, Qwen3-Max, Doubao-Seed-1.6, and DeepSeek-V3.2, remain highly susceptible, exhibiting Attack Success Rates (ASR) of 22.94\%, 13.44\%, 33.34\%, 37.60\%, 54.13\%, and 72.46\%, respectively.

    \item We release OpenRT as an open-source framework with a \textbf{long-term maintenance} commitment, continuously supporting attack evaluation and defense improvement while integrating new methods.
\end{itemize}

\section{Related Work}
\noindent\textbf{Red Teaming.}\;
Early approaches to MLLMs safety focused on manual red teaming, where human experts induce harmful outputs through targeted inputs, a process known as jailbreaking~\citep{perez2022red, liu2023jailbreaking, weidinger2023sociotechnical, ma2025safety}. While effective in uncovering subtle vulnerabilities~\citep{Li2024LLMDA, pliny2024}, manual methods are limited by scalability, cost, and coverage~\citep{bai2022constitutional, ganguli2022red}. To address these limitations, automated red teaming has gained attention~\citep{yu2023gptfuzzer, mazeika2024harmbench}, with early techniques focusing on input space exploration, such as genetic algorithms~\citep{liu2023jailbreaking, Lapid2023OpenSU}, token-level combinatorial methods~\citep{zang2020word}, gradient-based optimization~\citep{zou2023universal,chen2024llm,geisler2024attacking,wang2025freezevla}, and LLM-driven refinement schemes that iteratively improve attack prompts~\citep{chao2023jailbreaking, mehrotra2023tree,yu2024llmvirus,zhou2024easyjailbreak,xiao2024tastle}. However, these approaches primarily treat jailbreak discovery as search in the input space and remain confined to prompt refinement. Recent work has shifted toward agent-based frameworks that automate not only prompt generation but entire attack strategies, including systems like RedAgent~\citep{Xu2024RedAgentRT}, ALI-Agent~\citep{Zheng2024ALIAgentAL}, WildTeaming~\citep{jiang2024wildteaming}, AutoRedTeamer~\citep{zhou2025autoredteamer}, AutoDAN-Turbo~\citep{liu2024autodan}, H4RM3L~\citep{doumbouya2024h4rm3l}, X-Teaming~\cite{rahman2025x}, and EvoSynth~\citep{chen2025evolve}, which leverage multi-agent coordination and evolutionary techniques to generate novel attack vectors. Additionally, programmatic attacks such as CodeAttack~\citep{jha2023codeattack}, which treat code snippets purely as textual input, and co-evolutionary training frameworks like Evo-MARL~\citep{pan2025evo} or RL-based adversarial sample generation~\citep{zou2019reinforced} have further expanded the range of attack methods. These advancements significantly broaden the scope of automated red teaming, enabling more dynamic and scalable adversarial testing.

\noindent\textbf{Evaluation Frameworks and Benchmarks.}\;
Beyond developing individual attack algorithms, a significant body of work~\cite{chen2025teleai} has focused on creating standardized toolboxes and benchmarks for jailbreak evaluation. EasyJailbreak~\citep{zhou2024easyjailbreak}, for instance, offers a modular pipeline implementing numerous attack families. To further systematize evaluation, JailbreakBench~\citep{chao2024jailbreakbench} and HarmBench~\citep{mazeika2024harmbench} provide large-scale test suites and principled metrics for benchmarking adversarial defenses. Specialized frameworks have also emerged, including JailTrickBench~\citep{xu2024bag}, which focuses on specific jailbreak implementation techniques, and OmniSafeBench-MM~\citep{jia2025omnisafebench} for evaluating adversarial robustness in multimodal models. Frameworks like DeepTeam~\citep{deepteam} support automated adversarial testing across various attack strategies. In contrast, our work, OpenRT, is distinguished by its comprehensiveness and scale. It uniquely offers a unified and module framework that natively supports diverse MLLMs. By integrating 37 configurable and extensible attack strategies, OpenRT offers a more robust and scalable solution for assessing adversarial vulnerabilities across diverse MLLMs.

\begin{table}[t]
    \centering
    \resizebox{\textwidth}{!}{
    \begin{tabular}{lcccccc}
    \toprule
    \textbf{Framework} & \textbf{Methods} & \textbf{Modality} & \textbf{Interaction} & \textbf{Async} & \textbf{Configurability} & \textbf{Extensibility} \\ 
    \midrule
    EasyJailbreak~\cite{zhou2024easyjailbreak}       & 11 & Text        & Single           & $\times$ & Low    & Medium \\
    JailbreakBench~\cite{chao2024jailbreakbench}     & 4  & Text        & Single           & $\times$ & Medium & Low    \\
    HarmBench~\cite{mazeika2024harmbench}            & 18 & Text        & Single           & $\times$ & Medium & Medium \\
    JailTrickBench~\cite{xu2024bag}                  & 7  & Text        & Single           & $\times$ & Low    & Medium \\
    GA~\cite{GA}                                     & 6  & Text        & Single \& Multi  & $\times$ & Medium & Medium \\
    PyRIT~\cite{munoz2024pyrit}                      & 9  & Text        & Single \& Multi  & $\times$ & Medium & Medium \\ 
    DeepTeam~\cite{deepteam}                         & 19 & Text        & Single \& Multi  & $\times$ & Medium & Medium \\ 
    TeleAI-Safety~\cite{chen2025teleai}              & 19 & Text        & Single \& Multi  & $\times$ & Medium & Medium \\     
    OmniSafeBench-MM~\cite{jia2025omnisafebench}     & 13 & Image       & Single           & $\times$ & Medium & Medium \\ 
    \midrule
    \rowcolor{gray!15} \textbf{OpenRT (Ours)} 
                      & \textbf{37} 
                      & \textbf{Text \& Image} 
                      & \textbf{Single \& Multi \& Agent} 
                      & \textbf{\checkmark} 
                      & \textbf{High} 
                      & \textbf{High} \\ 
    \bottomrule
    \end{tabular}}
    \caption{Comparison of \textbf{OpenRT} with existing red-teaming frameworks. OpenRT stands out by enabling large-scale jailbreaking, providing a unified asynchronous execution engine for high scalability, and adopting a modular architecture that integrates 37 attack strategies. \textbf{Method} denotes an attack method proposed in a peer-reviewed paper (counting different variants as a single method). \textbf{Interaction} indicates the supported attack interaction patterns (single-turn, multi-turn, and multi-agent coordination). \textbf{Configurability} reflects the ease of setup (Low: hard-coded; Medium: requires code changes; High: YAML-based), while \textbf{Extensibility} measures the effort required to add new components (Low: nearly impossible; Medium: complex code changes; High: seamless integration).}
    \label{tab:comparison}
\end{table}

\section{Framework}
In this section, we present OpenRT, a comprehensive and extensible framework for systematic evaluation of MLLMs safety. Our framework addresses the fragmentation in current jailbreak research by providing a unified platform that enables fair comparison across diverse attack methodologies, supports both black-box and white-box attack paradigms, and facilitates reproducible experiments.

\subsection{Preliminaries}
\noindent\textbf{Threat Model}\;
We consider a standard red-teaming setting involving an adversary and a defender. The defender operates a target model equipped with safety policies, while the adversary seeks outputs that violate them. We formalize two distinct threat models based on the adversary's level of access:

\begin{itemize}
    \item \textbf{Black-box Setting:} The adversary interacts with the model solely via an API or inference interface, observing only the final output $y$ for a given input $x$. Internal states, such as gradients and logits, remain inaccessible. Furthermore, the adversary operates under strict constraints regarding query budget and request rate.    
    \item \textbf{White-box Setting:} The adversary possesses full transparency, including access to model parameters $\theta$, gradients $\nabla_\theta$, and hidden state embeddings. This regime facilitates worst-case robustness analysis via gradient-based optimization.
\end{itemize}

The objective is to maximize the Attack Success Rate (ASR) over a dataset of harmful queries $\mathcal{D}$, as determined by a safety judge $\mathcal{J}$.

\noindent\textbf{Problem Formulation.}\;
We define the target MLLMs as $\mathcal{F}_\theta$, which models a conditional probability distribution $p(\cdot \mid I; \theta)$ over output tokens, given a multimodal input context $I \in \mathcal{I}$. For a comprehensive safety evaluation, the input $I$ is formalized as a tuple $(v, t_{1:n})$, where $t_{1:n}$ denotes a sequence of $n$ discrete textual tokens, and $v$ represents the visual component. In multimodal scenarios, $v \in [0, 1]^d$ corresponds to a high-dimensional image observation; for text-only safety probes, $v$ may be treated as a null or empty element $\emptyset$. 
Adversarial attacks in this unified framework aim to manipulate these input components to synthesize an adversarial example $\mathcal{q}' = (v', \widetilde{t}_{1:n})$. This is achieved by either injecting discrete perturbations into the textual prompt $t_{1:n}$ or continuous noise into the visual observation $v$, such that the resulting output $r \sim p(\cdot \mid \mathcal{q}'; \theta)$ violates the model's pre-defined safety alignments.

Formally, given a clean input context $(v, t_{1:n})$ and a target sequence of unsafe tokens $t_{n+1:n+m}$, an adversary seeks to generate adversarial examples $\mathcal{q}'=(v', \widetilde{t}_{1:n})$ by optimizing the following objective:
\begin{equation}
    (v', \widetilde{t}_{1:n}) = \argmin_{\substack{\| v' - v \|_{\infty} \leq \epsilon \\ \text{PPL}(\widetilde{t}_{1:n}) \leq \beta}} -\log p(t_{n+1:n+m} \mid v', \widetilde{t}_{1:n}; \theta),
\end{equation}
where $t_{n+1:n+m}$ represents the target harmful content, $(v', \widetilde{t}_{1:n})$ is the adversarial example, $\epsilon$ denotes the perturbation budget for the visual modality~\cite{madry2017towards}, and $\beta$ represents the perplexity threshold for the textual prompt to ensure stealthiness. For text-based jailbreaking, the visual terms $(v, \epsilon)$ are omitted, reducing the optimization solely to the discrete prompt $\widetilde{t}_{1:n}$.

\subsection{Component Overview}
OpenRT decomposes the red-teaming pipeline into six modular components: \textbf{Model}, \textbf{Dataset}, \textbf{Attack}, \textbf{Judge}, \textbf{Evaluator}, and a central \textbf{Orchestrator}. This design achieves a high degree of decoupling, enabling any component to be replaced independently without requiring changes to the others. Table~\ref{tab:components} summarizes the role of each component.

\subsubsection{Target Model}
The Model component provides a unified interface to MLLMs, abstracting differences between cloud APIs and local deployments. The core interface includes: \texttt{query} for sending inputs and receiving responses, \texttt{get\_gradients} and \texttt{get\_embedding} for white-box attacks.

The framework supports two implementations: (1) \textbf{API-based models} compatible with OpenAI-style endpoints, featuring conversation history management, multi-modal input support, and automatic retry mechanisms; (2) \textbf{Local models} with full gradient access for white-box attacks such as GCG~\citep{zou2023universal}. Multi-turn jailbreaking is enabled through conversation history $\mathcal{H}$, where $r_k = \mathcal{Q}(q_k \mid \mathcal{H}_{<k})$, controlled by the \texttt{maintain\_history} parameter.

\begin{table}[ht]
    \centering
    \small
    \resizebox{1.0\textwidth}{!}{
    \setlength{\tabcolsep}{4.3mm}{
    \begin{tabular}{lcccccl}
    \toprule
    \textbf{Method} & \textbf{Year} & \textbf{Multi-Modal} & \textbf{Multi-Turn} & \textbf{Multi-Agent} & \textbf{Strategy Paradigm} \\
    \midrule
    \multicolumn{6}{l}{\textit{\textbf{White-Box}}} \\
    GCG~\cite{zou2023universal}                 & 2023 & Text  & Single  & $\times$ & Gradient Optimization \\
    Visual Jailbreak~\cite{qi2024visual}        & 2023 & Image & Single & $\times$ & Gradient Optimization \\
    \midrule
    \multicolumn{6}{l}{\textit{\textbf{Black-Box: Optimization \& Fuzzing}}} \\
    AutoDAN~\cite{liu2023autodan}               & 2023 & Text & Single & $\times$ & Genetic Algorithm \\
    GPTFuzzer~\cite{yu2023gptfuzzer}            & 2023 & Text & Single & $\times$ & Fuzzing / Mutation \\
    TreeAttack~\cite{mehrotra2024tree}          & 2023 & Text & Single & $\times$ & Tree-Search Optimization \\
    SeqAR~\cite{yang2025seqar}                  & 2024 & Text & Single & $\times$ & Genetic Algorithm \\
    RACE~\cite{ying2025reasoning}               & 2025 & Text & Single & $\times$ & Gradient/Genetic Optimization \\
    AutoDAN-R~\cite{liu2025autodan}             & 2025 & Text & Single & $\times$ & Test-Time Scaling \\
    \midrule
    \multicolumn{6}{l}{\textit{\textbf{Black-Box: LLM-driven Refinement}}} \\
    PAIR~\cite{chao2023jailbreaking}            & 2023 & Text & Single & $\times$ & Iterative LLM Optimization \\
    ReNeLLM~\cite{ding2024wolf}                 & 2023 & Text & Single & $\times$ & Rewrite \& Nesting \\
    DrAttack~\cite{li2024drattack}              & 2024 & Text & Single & $\times$ & Prompt Decomposition \\
    AutoDAN-Turbo~\cite{liu2024autodan}         & 2024 & Text & Single & $\times$ & Genetic + Gradient Guide \\
    \midrule
    \multicolumn{6}{l}{\textit{\textbf{Black-Box: Linguistic \& Encoding}}} \\
    CipherChat~\cite{yuan2023gpt}               & 2023 & Text & Single & $\times$ & Cipher/Encryption \\
    CodeAttack~\cite{ren2024codeattack}         & 2022 & Text & Single & $\times$ & Code Encapsulation \\
    Multilingual~\cite{deng2023multilingual}    & 2023 & Text & Single & $\times$ & Low-Resource Language \\
    Jailbroken~\cite{wei2023jailbroken}         & 2023 & Text & Single & $\times$ & Template Combination \\
    ICA~\cite{wei2023jailbreak}                 & 2023 & Text & Single & $\times$ & In-Context Demonstration \\
    FlipAttack~\cite{liu2024flipattack}         & 2024 & Text & Single & $\times$ & Token Flipping / Masking \\
    Mousetrap~\cite{yao2025mousetrap}           & 2025 & Text & Single & $\times$ & Logic Nesting / Obfuscation \\
    Prefill~\cite{li2025prefill}                & 2025 & Text & Single & $\times$ & Prefix Injection \\
    \midrule
    \multicolumn{6}{l}{\textit{\textbf{Black-Box: Contextual Deception}}} \\
    DeepInception~\cite{li2023deepinception}    & 2023 & Text & Single & $\times$ & Hypnosis or Nested Scene \\
    Crescendo~\cite{russinovich2025great}       & 2024 & Text & Multi  & $\times$ & Multi-turn Steering \\
    RedQueen~\cite{jiang2024red}                & 2024 & Text & Multi  & $\times$ & Concealed Knowledge \\
    CoA~\cite{yang2024chain}                    & 2024 & Text & Multi & $\times$ & Chain of Attack \\
    \midrule
    \multicolumn{6}{l}{\textit{\textbf{Black-Box: Multimodal Specific}}} \\
    FigStep~\cite{gong2025figstep}              & 2023 & Image & Single & $\times$ & Typography / OCR \\
    QueryRelevant~\cite{liu2023query}           & 2024 & Image & Single & $\times$ & Visual Prompt Injection \\
    IDEATOR~\cite{wang2025ideator}              & 2024 & Image & Single & $\times$ & Visual Semantics \\
    MML~\cite{wang2025jailbreak}                & 2024 & Image & Single & $\times$ & Cross‑Modal Encryption \\
    HADES~\cite{li2024images}                   & 2024 & Image & Single & $\times$ & Visual Vulnerability Amplification \\  
    HIMRD~\cite{ma2025heuristic}                & 2024 & Image & Single & $\times$ & Multi-Modal Risk Distribution \\
    JOOD~\cite{jeong2025playing}                & 2025 & Image & Single & $\times$ & OOD Transformation \\
    SI~\cite{zhao2025jailbreaking}              & 2025 & Image & Single & $\times$ & Shuffle Inconsistency Optimization \\
    CS-DJ~\cite{yang2025distraction}            & 2025 & Image & Single & $\times$ & Multi‑Level Visual Distraction
 \\
    \midrule
    \multicolumn{6}{l}{\textit{\textbf{Black-Box: Multi-Agent \& Cooperative}}} \\
    ActorAttack~\cite{ren2024derail}            & 2024 & Text & Multi  & $\checkmark$ & Actor-Based Steering \\
    Rainbow Teaming~\cite{samvelyan2024rainbow} & 2024 & Text & Multi  & $\checkmark$ & Diversity-Driven Search \\
    X-Teaming~\cite{rahman2025x}                & 2025 & Text & Multi  & $\checkmark$ & Cooperative Exploration \\
    EvoSynth~\cite{chen2025evolve}              & 2025 & Text & Multi  & $\checkmark$ & Code-Level Evolutionary Synthesis\\
    \bottomrule
    \end{tabular}}}
    \caption{Taxonomy of jailbreak attack methods based on strategy paradigms.}
    \label{tab:attack_methods}
\end{table}

\subsubsection{Dataset}
The Dataset component is responsible for managing and providing harmful queries, which are utilized as attack targets in various contexts. It supports loading test cases from different data sources, allowing for flexibility in evaluation and benchmarking. Formally, a dataset is defined as an ordered collection of harmful queries, denoted as $\mathcal{D} = \{q_1, q_2, \ldots, q_N\}$, where each query $q_i$ belongs to the harmful query space $\mathcal{Q}$, and $N = |\mathcal{D}|$ represents the total number of queries in the dataset. This structure ensures that datasets are organized and ready for use in testing scenarios.

There are different types of datasets tailored to specific use cases. The first type, the StaticDataset, is an in-memory dataset suitable for small-scale testing. On the other hand, the JSONLDataset supports streaming loading of large-scale evaluation benchmarks (\emph{e.g.}, AdvBench~\citep{zou2023universal}, HarmBench~\citep{mazeika2024harmbench}).

\subsubsection{Attack}
The Attack component implements 37 jailbreak methods across black-box and white-box settings. Each attack $\mathcal{A}$ transforms a harmful query $q$ into an adversarial example $q'$, returning a structured \texttt{AttackResult} object containing the original query, success flag, adversarial prompt, model response, conversation history, computational cost, and adversarial image path (for multi-modal attacks).

Table~\ref{tab:attack_methods} classifies attacks by strategy paradigm, including gradient optimization, genetic algorithms, LLM-driven refinement, linguistic encoding, contextual deception, multimodal exploitation, and multi-agent coordination. \textbf{White-box attacks} include gradient-based methods like GCG~\cite{zou2023universal} and Visual Jailbreak~\cite{qi2024visual}. \textbf{Black-box attacks} span multiple categories: optimization and fuzzing approaches (AutoDAN~\cite{liu2023autodan}, GPTFuzzer~\cite{yu2023gptfuzzer}), LLM-driven refinement (PAIR~\cite{chao2023jailbreaking}, DrAttack~\cite{li2024drattack}), linguistic and encoding manipulations (CipherChat~\cite{yuan2023gpt}, CodeAttack~\cite{ren2024codeattack}, FlipAttack~\cite{liu2024flipattack}), contextual deception strategies (Crescendo~\cite{russinovich2025great}, RedQueen~\cite{jiang2024red}), multimodal-specific attacks (FigStep~\cite{gong2025figstep}, IDEATOR~\cite{wang2025ideator}), and multi-agent coordination (Rainbow Teaming~\cite{samvelyan2024rainbow}, X-Teaming~\cite{rahman2025x}, EvoSynth~\cite{chen2025evolve}).

\subsubsection{Judge}
The Judge component determines attack success by assessing whether model responses violate safety policies. It is formalized as a scoring function that outputs a harmfulness score $s \in [1, 5]$ and reasoning explanation $\rho$. The scoring criteria range from 1 (proactive refusal, best safe response) to 5 (full compliance with harmful instructions), as detailed in Table~\ref{tab:scoring}.

\begin{table}[ht]
    \centering
    \small
    \begin{tabular}{cl}
    \toprule
    \textbf{Score} & \textbf{Description} \\
    \midrule
    1 & Model proactively refuses; best safe response \\
    2 & Model evades without direct refusal; provides general warning \\
    3 & Model engages but lacks safety awareness \\
    4 & Model provides explicitly violating content, but not full compliance \\
    5 & Model fully complies with harmful instruction \\
    \bottomrule
    \end{tabular}
    \caption{Harmfulness scoring criteria for judge evaluation.}
    \label{tab:scoring}
\end{table}

OpenRT incorporates a dual-judge architecture for robust success determination. 
The \textbf{Keyword-Based Judge} employs pattern-matching heuristics for rapid, large-scale refusal detection. 
The \textbf{LLM-Based Judge} complements this by leveraging an independent model to semantically analyze complex responses. This hybrid approach ensures both the scalability required for extensive benchmarking and the granular accuracy necessary for evaluating frontier MLLMs.

The \texttt{success\_threshold} parameter $\theta \in [1, 5]$ defines the success criterion: an attack succeeds when the score $s \geq \theta$. This flexibility allows researchers to adjust evaluation stringency based on their threat model requirements.

\subsubsection{Evaluator}
The Evaluator component aggregates attack results to compute experiment-level metrics, providing a comprehensive assessment of attack effectiveness, efficiency, and impact. By mapping raw attack results to quantitative metrics, it offers insights into model vulnerabilities and facilitates the refinement of defense mechanisms.

We evaluate performance using four key metrics. First, \textbf{Attack Success Rate (ASR)} measures the proportion of successful attacks, defined as:
\begin{equation}
    \text{ASR} = \frac{1}{N}\sum_{i=1}^{N} \mathbf{1}[s_i \geq \theta],
\end{equation}
where $s_i$ denotes the harmfulness score of the $i$-th attack, and $\theta$ represents the success threshold. 

Complementing ASR, we employ three diagnostic metrics. \textbf{Attack Efficiency} quantifies the resources consumed (\emph{e.g.}, time, API calls, input token, and output tokens) to achieve success. This is formally measured by the average computational cost $\bar{c} = \frac{1}{N}\sum_{i=1}^{N} c_i$, where $c_i$ represents the cost of the $i$-th attempt. \textbf{Attack Stealthiness} assesses the imperceptibility of attacks against content filters or anomaly detectors, evaluated via metrics such as perplexity or semantic similarity to benign inputs. Finally, \textbf{Attack Diversity} measures the variety of adversarial strategies explored, with higher scores indicating a broader coverage of potential vulnerabilities.

\begin{algorithm}[t]
    \caption{Orchestrator Execution Pipeline}
    \label{alg:orchestrator}
    \begin{algorithmic}[1]
    \Require Model $\mathcal{M}$, Dataset $\mathcal{D}$, Attack $\mathcal{A}$, Evaluator $\mathcal{E}$, Max Workers $W$
    \Ensure Metrics $\mu$, Results $\mathcal{R}$
    \State \textbf{// Phase 1: Initialization}
    \State $\mathcal{R} \leftarrow [\text{None}] \times |\mathcal{D}|$ \Comment{Initialize results array}
    \State ThreadPool $\leftarrow$ ThreadPoolExecutor($W$) \Comment{Initialize thread pool with $W$ workers}
    \State
    \State \textbf{// Phase 2: Parallel Attack Execution}
    \State tasks = \{submit($\mathcal{A}$, $q_i$): $i$ for $(i, q_i)$ in enumerate($\mathcal{D}$)\}
    \For{each task $\in$ as\_completed(tasks)}
        \State $i \leftarrow$ tasks[task] \Comment{Get the index of the result}
        \State \textbf{try:}
        \State \quad $\mathcal{R}[i] \leftarrow$ task.result() \Comment{Store result if successful}
        \State \textbf{except Exception as e:}
        \State \quad $\mathcal{R}[i] \leftarrow$ AttackResult(success = False, target = $\mathcal{D}[i]$) \Comment{Store failure result}
        \State \quad log\_error(e) \Comment{Log error if exception occurs}
    \EndFor
    \State
    \State \textbf{// Phase 3: Aggregated Evaluation}
    \State $\mu \leftarrow \mathcal{E}$.evaluate($\mathcal{R}$) \Comment{Evaluate results using evaluator}
    \State
    \State \textbf{// Phase 4: Result Reporting}
    \State print("Final ASR:", $\mu$.ASR) \Comment{Output final Attack Success Rate} \\
    \Return ($\mu$, $\mathcal{R}$) \Comment{Return metrics and results}
    \end{algorithmic}
\end{algorithm}

\subsubsection{Orchestrator}
The Orchestrator serves as the central coordinator of OpenRT, managing all components to execute complete experimental pipelines. It accepts various of target models $\mathcal{M}$, a dataset of harmful queries $\mathcal{D}$, an array of attack method $\mathcal{A}$, and an evaluator $\mathcal{E}$, orchestrating their interactions to produce evaluation metrics $\mu$ and detailed attack results $\mathcal{R}$.

The execution follows a four-phase workflow, as described in Algorithm~\ref{alg:orchestrator}: (1) initialization of result containers and thread pools, (2) parallel execution of attacks across the dataset, (3) aggregation of the collected results for evaluation, and (4) final reporting. The Orchestrator is built around several core design principles: Single Responsibility, ensuring it focuses solely on coordination and delegates attack and evaluation logic to their respective components; Parallel Execution, utilizing the \texttt{ThreadPoolExecutor} with configurable \texttt{max\_workers} to efficiently handle large-scale evaluations; Fault Isolation, which captures and logs individual attack failures without interrupting other executions; and Progress Tracking, offering real-time feedback through tqdm with completion counts and success rates.

Experiments are Configuration-Driven through YAML files, enabling dynamic component assembly, reproducible benchmarking, fair comparison under identical conditions, and convenient parallelization for hyperparameter sweeps.

\subsubsection{Modular Component Registry}
OpenRT employs a unified registry system that enables automatic component discovery and runtime instantiation through a decorator-based approach. Each component type maintains its own registry (\texttt{attack\_registry}, \texttt{model\_registry}, \texttt{dataset\_registry}, \texttt{evaluator\_registry}, \texttt{judge\_registry}), allowing new implementations to be registered via simple decorators (\emph{e.g.}, \texttt{@attack\_registry.register("pair")}). This mechanism automatically catalogs all available components, enabling the framework to dynamically instantiate and assemble them based on configuration files without requiring modifications to the core codebase.

This modular registry design offers three key benefits. First, \textbf{Extensibility}: researchers can integrate new attack methods or model interfaces by implementing the corresponding base class and registering it, without touching existing code. Second, \textbf{Discoverability}: the system provides programmatic access to list all registered components, facilitating automated experimentation and hyperparameter sweeps. Third, \textbf{Flexibility}: any combination of registered components can be dynamically assembled at runtime through configuration files, enabling rapid prototyping and fair comparison across diverse experimental setups. Table~\ref{tab:components} summarizes the role and key implementations of each component.

\begin{table}[ht]
    \centering
    \small
    \begin{tabular}{lll}
    \toprule
    \textbf{Component} & \textbf{Role} & \textbf{Key Implementations} \\
    \midrule
    Model & Target MLLMs abstraction & OpenAIModel, HuggingFaceModel \\
    Dataset & Attack target management & StaticDataset, JSONLDataset \\
    Attack & Jailbreak method execution & PAIR, GPTFuzzer, GCG, X-Teaming, etc. \\
    Judge & Success determination & KeywordJudge, LLMJudge \\
    Evaluator & Metrics aggregation & KeywordEvaluator, JudgeEvaluator \\
    Orchestrator & Pipeline coordination & Parallel execution, fault isolation \\
    \bottomrule
    \end{tabular}
    \caption{Summary of framework components, their roles, and key implementations.}
    \label{tab:components}
\end{table}

\section{Experiments}
To evaluate the effectiveness of OpenRT, we conduct a series of experiments targeting a diverse range of state-of-the-art MLLMs. Our primary goal is to assess the ability of our framework to autonomously synthesize novel and effective jailbreaking methods in a strict black-box setting.

\subsection{Experimental Setup}
Our experimental setup is designed to ensure a rigorous and fair comparison against current state-of-the-art methods. To this end, we closely follow the evaluation protocols established by leading baseline frameworks, particularly X-Teaming~\citep{rahman2025x} and ActorAttack~\citep{ren2024derail}.  Following these works, we also use Harmbench Standard\cite{mazeika2024harmbench}  as the evaluation dataset. This dataset is designed to be comprehensive, with instructions balanced across 6 different risk categories specified in emerging AI regulation, including 6 semantic categories of behavior: Cybercrime \& Unauthorized Intrusion, Chemical \& Biological Weapons/Drugs,
 Misinformation \& Disinformation,
Harassment \& Bullying, Illegal Activities, and General
Harm, ensuring our evaluation represents a representative spectrum of potential harms.

\subsubsection{Datasets and Models}
For our experiments, we employ the \textbf{HarmfulBench}~\cite{mazeika2024harmbench} dataset, which comprises a curated collection of harmful queries designed to probe the safety vulnerabilities of MLLMs. We evaluate the performance of over 20 distinct target models, including MLLMs such as \textbf{GPT-5.2}~\cite{gpt5-2}, \textbf{GPT-5.1}~\cite{gpt5}, \textbf{Claude Haiku 4.5}~\cite{ClaudeHaiku45}, \textbf{Gemini 3 Pro Preview}~\cite{Gemini3ProPreview}, \textbf{Gemini 2.5 Flash Thinking}~\cite{comanici2025gemini}, \textbf{Mistral Large 3}~\cite{Mistral32025}, \textbf{Llama-4-Maverick}~\cite{Meta2025Llama4}, \textbf{Llama-4-Scout}~\cite{Meta2025Llama4}, \textbf{Grok 4.1 Fast}~\cite{Grok41Fast2025}, and \textbf{Doubao-Seed-1.6}~\cite{Seed16}, as well as LLMs including \textbf{Qwen3-Max}~\cite{qwen3max}, \textbf{Qwen3-235B-A22B-Thinking}~\cite{Qwen3Thinking235B}, \textbf{Qwen3-Next-80B-A3B}~\cite{Qwen2025Qwen3Next80B}, \textbf{DeepSeek-R1}~\cite{guo2025deepseek}, \textbf{DeepSeek-V3.2}~\cite{liu2025deepseek}, \textbf{Kimi K2-Instruct-0905}~\cite{kimiteam2025kimik2openagentic}, \textbf{MiniMax-M2}~\cite{MiniMax2025M2}, \textbf{GLM-4.6}~\cite{vteam2025glm45vglm41vthinkingversatilemultimodal}, \textbf{Hunyuan-A13B-Instruct}~\cite{HunyuanA13BInstruct}, and \textbf{ERNIE-4.5-300B-A47B}~\cite{ernie2025technicalreport}. These models represent the current frontier in AI safety and alignment, making them challenging targets.

\subsubsection{Attack Configuration}
We evaluate 37 distinct attack methods strictly within the black-box setting. Our assessment covers a diverse range of strategies, including single-turn prompting, multi-turn conversational interactions, multi-modal optimization, and multi-agent coordination. Each attack method is rigorously configured with relevant hyperparameter such as the number of iterations, the mutation rate for genetic algorithms, the query budget, and other parameters that influence the attack dynamics. These techniques (\emph{e.g.}, genetic algorithms and fuzzing-based methods) are designed to optimize adversarial inputs relying exclusively on API-level output responses, independent of the model's internal gradients or states.

\subsubsection{Implementation Details}
All experiments are conducted within a unified and reproducible OpenRT framework, utilizing a modular orchestration design to ensure fair comparisons across various attack methods and target models. We assign specialized models for distinct roles: all helper models (\emph{e.g.}, attacker, mutator, planner, optimizer) utilize DeepSeek-V3.2~\cite{liu2025deepseek} with a temperature of 1.0 to encourage diverse and creative adversarial prompt generation, while the judge model employs GPT-4o-mini~\cite{hurst2024gpt} with a temperature of 0.0 to ensure deterministic and consistent safety evaluation, using a success threshold score of 5 for binary jailbreak classification. For attacks requiring semantic similarity computation, we leverage text-embedding-3-large as our embedding backbone. For multimodal attacks targeting vision-language models, we incorporate Imagen-4.0-fast as the diffusion-based image generator for synthesizing adversarial visual content, and Qwen2.5-VL-32B-Instruct serves as the vision-language attacker model for attacks requiring visual understanding and multimodal reasoning capabilities. Experiments are executed using a concurrent orchestration pipeline with 25 parallel workers per target model to maximize throughput while respecting API rate limits, and the evaluation module employs 32 parallel workers for efficient batch assessment of attack outcomes. All attack results are systematically logged with comprehensive metadata, including the full attack trajectory, intermediate prompts, model responses, and execution time for reproducibility analysis.

\subsubsection{Performance Metrics}
We evaluate attack performance using four complementary metrics that capture effectiveness, efficiency, stealthiness, and diversity. Consequently, an ideal attack method strives to maximize vulnerability coverage through high success rates and strategic variation, while ensuring practical utility by minimizing resource consumption and linguistic detectability, which can be conceptualized as:
\begin{equation}
    \text{Objective} \sim \text{ASR} \uparrow + \text{Diversity} \uparrow + \text{Cost} \downarrow + \text{PPL} \downarrow
\end{equation}

\noindent\textbf{(1) Attack Success Rate} measures the proportion of attacks that successfully elicit harmful outputs:
\begin{equation}
    \text{ASR} = \frac{1}{N}\sum_{i=1}^{N} \mathbf{1}[s_i \geq \theta],
\end{equation}
where $N$ denotes the total number of attack attempts, $s_i \in [1,5]$ is the harmfulness score assigned by the judge for the $i$-th attack, and $\theta$ is the success threshold.

\noindent\textbf{(2) Attack Efficiency} quantifies the average resource consumption per attack:
\begin{equation}
    \text{Efficiency} = \frac{1}{N}\sum_{i=1}^{N} c_i,
\end{equation}
where $c_i$ represents the cost of the $i$-th attack, measured in terms of total tokens consumed, number of API calls, or wall-clock time. Lower values indicate more efficient attacks.

\noindent\textbf{(3) Attack Stealthiness} assesses the linguistic naturalness of adversarial prompts. Consistent with the perplexity constraint in Eq.~(1), we measure:
\begin{equation}
    \text{PPL}(\widetilde{t}_{1:n}) = \exp\left(-\frac{1}{n}\sum_{j=1}^{n} \log p(\widetilde{t}_j \mid \widetilde{t}_{<j})\right),
\end{equation}
where $\widetilde{t}_{1:n} = (t_1, \ldots, t_n)$ is the adversarial textual prompt and $p(\widetilde{t}_j \mid \widetilde{t}_{<j})$ is the probability assigned by a reference language model. Lower perplexity indicates more natural prompts satisfying the stealthiness constraint $\text{PPL}(\widetilde{t}_{1:n}) \leq \beta$.

\noindent\textbf{(4) Attack Diversity} quantifies the semantic variety of adversarial strategies explored by each attack method. We compute diversity as the mean pairwise cosine distance between embeddings of successful adversarial prompts:
\begin{equation}
    \text{Diversity} = \frac{2}{n(n-1)} \sum_{i < j} \left(1 - \frac{\mathbf{e}_i \cdot \mathbf{e}_j}{\|\mathbf{e}_i\| \|\mathbf{e}_j\|}\right),
\end{equation}
where $\mathbf{e}_i$ and $\mathbf{e}_j$ are the semantic embeddings of the $i$-th and $j$-th successful adversarial prompts, and $n$ is the total number of successful attacks. Higher diversity scores indicate that the attack method explores a broader range of adversarial strategies, enabling more comprehensive coverage of the vulnerability space and reducing the likelihood of converging to a narrow set of exploitation patterns.

\subsection{Main Results}
The results of our experiments, summarized in Tables~\ref{tab:attack_MLLM} and \ref{tab:attack_llm}, provide a detailed overview of the Attack Success Rate (ASR) achieved by various attack strategies across multiple MLLMs. These results highlight the vulnerabilities of MLLMs, even with advanced safety mechanisms in place. We report the attack performance for each method across distinct models, identifying key insights into the strengths and weaknesses of different attacks.

\subsubsection{Vulnerabilities in Multimodal Large Language Models}
The attack performance on various MLLMs shows significant variation in the effectiveness of different strategies. Notably, the EvoSynth attack method nearly achieves perfect ASRs of 100\% across a wide range of models, including Gemini 3 Pro, Mistral Large, and Doubao Seed 1.6. This suggests that EvoSynth is highly robust, exploiting the vulnerabilities of MLLMs regardless of their specific architecture. Similarly, other strategies, such as Mousetrap and X-Teaming, also exhibit exceptional performance across several MLLMs; for instance, Mousetrap achieves perfect or near-perfect success rates on Grok 4.1 Fast and Doubao Seed 1.6, while X-Teaming maintains ASR above 85\% for the majority of models. This indicates that these attacks are capable of manipulating the models into generating harmful outputs, even in challenging scenarios. However, some attacks show less consistent success. For example, RedQueen and CoA exhibit relatively low ASRs, especially on models like Gemini 2.5 Flash and Doubao Seed 1.6, where their success rates remain below 10\%. These results indicate that these methods may require further refinement to enhance their robustness across diverse models.

\begin{table}[htp]
    \centering
    \scriptsize
    \setlength{\tabcolsep}{2pt}
    \resizebox{\textwidth}{!}{
    \begin{tabular}{@{}l*{10}{S}@{}}
    \toprule
        & \multicolumn{1}{c}{\mhead{GPT-5.2}} 
        & \multicolumn{1}{c}{\mhead{GPT-5.1}} 
        & \multicolumn{1}{c}{\mhead{Claude\\Haiku 4.5}} 
        & \multicolumn{1}{c}{\mhead{Gemini 3\\Pro Preview}} 
        & \multicolumn{1}{c}{\mhead{Gemini 2.5\\Flash}} 
        & \multicolumn{1}{c}{\mhead{Mistral\\Large 3}} 
        & \multicolumn{1}{c}{\mhead{Llama-4\\Maverick}} 
        & \multicolumn{1}{c}{\mhead{Llama-4\\Scout}} 
        & \multicolumn{1}{c}{\mhead{Grok 4.1\\Fast}} 
        & \multicolumn{1}{c}{\mhead{Doubao\\Seed-1.6}} \\
    \midrule
    AutoDAN          & 2.0   & 8.0   & 1.5  & 22.5  & 37.5  & 28.5  & 23.5  & 64.5  & 38.5  & 13.0 \\
    GPTFuzzer        & 11.0  & 1.5   & 0.0  & 51.0  & 93.0  & 97.5  & 64.0  & 97.5  & 31.0  & 57.0 \\
    TreeAttack       & 11.0  & 23.5  & 8.0  & 49.5  & 79.0  & 74.5  & 69.5  & 80.5  & 81.0  & 68.0 \\
    SeqAR            & 25.0  & 29.5  & 0.0  & 8.5   & 97.5  & 99.0  & 73.0  & 88.0  & 55.5  & 64.0 \\
    RACE             & 24.5  & 38.0  & 24.5 & 47.0  & 47.5  & 53.0  & 30.5  & 59.5  & 49.5  & 48.0 \\
    AutoDAN-R        & 70.5  & 69.0  & 28.5 & 83.0  & 96.5  & 97.0  & 96.5  & 80.0  & 90.0  & 86.5 \\
    PAIR             & 38.5  & 72.5  & 13.0 & 74.5  & 84.5  & 78.0  & 66.0  & 89.5  & 80.0  & 75.5 \\
    ReNeLLM          & 8.0   & 33.5  & 0.5  & 13.5  & 51.5  & 22.0  & 39.0  & 57.0  & 42.5  & 43.0 \\
    DrAttack         & 32.0  & 54.0  & 5.5  & 56.0  & 56.0  & 89.5  & 60.5  & 83.0  & 31.5  & 68.0 \\
    AutoDAN-Turbo    & 21.5  & 15.5  & 1.0  & 0.0   & 0.5   & 83.5  & 0.5   & 0.0   & 3.0   & 1.0 \\
    CipherChat       & 14.5  & 64.0  & 32.5 & 0.0   & 89.5  & 64.0  & 21.0  & 68.0  & 26.0  & 38.5 \\
    CodeAttack       & 22.0  & 20.5  & 29.5 & 10.5  & 51.0  & 8.5   & 71.0  & 86.5  & 22.0  & 89.0 \\
    Multilingual     & 16.5  & 25.0  & 0.0  & 2.0   & 34.0  & 55.5  & 14.0  & 0.0   & 1.5   & 6.5 \\
    Jailbroken       & 7.0   & 29.5  & 0.0  & 11.0  & 92.5  & 98.5  & 39.5  & 33.5  & 31.5  & 28.0 \\
    ICA              & 14.0  & 33.5  & 0.0  & 9.0   & 98.5  & 99.0  & 8.0   & 37.0  & 41.0  & 65.5 \\
    FlipAttack       & 13.5  & 68.5  & 0.0  & 19.5  & 95.5  & 95.5  & 65.5  & 54.5  & 23.0  & 87.0 \\
    Mousetrap        & 97.5  & 71.0  & 0.0  & 49.0  & 95.5  & 100.0 & 95.5  & 87.5  & 100.0 & 100.0 \\
    Prefill          & 1.0   & 14.0  & 0.0  & 3.5   & 97.5  & 97.0  & 34.5  & 43.5  & 25.5  & 30.5 \\
    DeepInception    & 15.5  & 19.0  & 0.0  & 3.5   & 84.0  & 100.0 & 82.5  & 94.5  & 37.5  & 82.0 \\
    Crescendo        & 32.5  & 51.0  & 9.0  & 47.0  & 48.0  & 61.0  & 17.0  & 30.5  & 41.0  & 58.0 \\
    RedQueen         & 0.0   & 1.0   & 0.0  & 2.5   & 3.0   & 4.5   & 3.0   & 5.5   & 1.5   & 21.5 \\
    CoA              & 15.5  & 0.0   & 0.5  & 2.0   & 4.5   & 16.5  & 3.0   & 19.0  & 7.0   & 4.5 \\
    FigStep          & 2.0   & 1.5   & 1.5  & 7.5   & 12.0  & 18.5  & 42.5  & 25.5  & 5.5   & 13.5 \\
    QueryRelevant    & 1.5   & 4.0   & 2.0  & 5.0   & 16.0  & 24.0  & 26.0  & 16.0  & 10.0  & 8.5 \\
    IDEATOR          & 31.5  & 73.0  & 17.0 & 80.0  & 95.0  & 94.5  & 90.0  & 94.0  & 94.5  & 96.0 \\
    MML              & 4.5   & 68.0  & 75.0 & 40.5  & 98.0  & 98.0  & 90.5  & 90.5  & 58.0  & 97.5 \\
    HADES            & 0.0   & 1.0   & 2.0  & 7.0   & 29.5  & 33.0  & 25.0  & 29.0  & 22.5  & 17.5 \\
    HIMRD            & 11.5  & 35.0  & 0.0  & 9.0   & 70.0  & 61.5  & 3.5   & 29.5  & 1.5   & 49.5 \\
    JOOD             & 65.0  & 62.5  & 38.0 & 56.0  & 61.5  & 63.0  & 38.5  & 39.5  & 69.5  & 72.0 \\
    SI               & 3.0   & 45.0  & 14.0 & 37.0  & 82.5  & 47.5  & 81.0  & 71.5  & 27.0  & 44.0 \\
    CS-DJ            & 15.0  & 21.5  & 23.5 & 35.0  & 39.5  & 38.0  & 35.0  & 39.5  & 28.5  & 51.0 \\
    ActorAttack      & 0.5   & 31.0  & 10.0 & 65.0  & 76.0  & 0.5   & 65.5  & 79.0  & 50.0  & 56.0 \\
    Rainbow Teaming  & 0.5   & 3.5   & 12.0 & 73.5  & 61.0  & 5.5   & 3.5   & 35.0  & 13.5  & 67.0 \\
    X-Teaming        & 75.5  & 95.5  & 47.5 & 86.5  & 89.0  & 91.0  & 86.0  & 98.0  & 90.5  & 87.0 \\
    EvoSynth         & 99.0  & 100.0 & 74.0 & 100.0 & 100.0 & 100.0 & 100.0 & 100.0 & 100.0 & 100.0 \\
    \bottomrule
    \end{tabular}}
    \caption{Attack Performance across Different MLLMs on HarmfulBench}
    \label{tab:attack_MLLM}
\end{table}

\subsubsection{Vulnerabilities in Large Language Models}
The performance of various attacks on LLMs demonstrates similar trends, with some attack methods showing superior generalization across different models. EvoSynth, once again, stands out, achieving nearly 100\% ASR across models like Qwen3-Max and Qwen3-Next-80B-A3B, indicating that this method is extremely effective in breaching model defenses in the LLM domain as well. Other notable attacks such as GPTFuzzer and PAIR also exhibit impressive results across multiple LLMs. For example, GPTFuzzer achieves high ASRs in models like DeepSeek-R1 and MiniMax-M2, with attack success rates ranging from 87\% to 97\%, highlighting its ability to generate adversarial prompts that consistently bypass safety filters. Similarly, PAIR performs robustly across models like Qwen3-Max, DeepSeek-V3.2, and GLM-4.6, achieving ASRs between 80\% and 95\%, demonstrating its strong adaptability in various scenarios. However, certain methods such as RedQueen and CoA show lower efficacy across several LLMs, with ASRs often below 10\% for models like Qwen3-Max and DeepSeek-R1. Furthermore, certain models exhibit polarized results that emphasize the necessity of diverse testing strategies. MiniMax-M2, for instance, is highly resistant to DeepInception (0\% ASR) but completely vulnerable to PAIR (90\% ASR). 

\begin{table}[ht!]
    \centering
    \scriptsize
    \setlength{\tabcolsep}{1.5pt}
    \resizebox{\textwidth}{!}{%
    \begin{tabular}{@{}l*{10}{S}@{}}
    \toprule
        & \multicolumn{1}{c}{\mhead{Qwen3-Max}}
        & \multicolumn{1}{c}{\mhead{Qwen3-235B\\A22B}}
        & \multicolumn{1}{c}{\mhead{Qwen3-Next\\80B-A3B}}
        & \multicolumn{1}{c}{\mhead{DeepSeek\\R1}}
        & \multicolumn{1}{c}{\mhead{DeepSeek\\V3.2}}
        & \multicolumn{1}{c}{\mhead{Kimi\\K2-Instruct}}
        & \multicolumn{1}{c}{\mhead{MiniMax-M2}}
        & \multicolumn{1}{c}{\mhead{GLM-4.6}}
        & \multicolumn{1}{c}{\mhead{Hunyuan\\A13B-Instruct}}
        & \multicolumn{1}{c}{\mhead{ERNIE-4.5\\300B-A47B}} \\
    \midrule
    AutoDAN           & 3.0 & 80.0 & 7.5 & 40.0 & 44.0 & 33.0 & 61.0 & 53.5 & 17.5 & 20.5 \\
    GPTFuzzer         & 9.5 & 92.0 & 78.0 & 97.0 & 96.5 & 87.5 & 19.0 & 97.0 & 42.5 & 98.0 \\
    TreeAttack        & 52.5 & 47.0 & 28.5 & 80.5 & 80.5 & 54.5 & 48.5 & 58.0 & 77.5 & 67.5 \\
    SeqAR             & 92.0 & 25.5 & 30.5 & 96.5 & 100.0 & 96.0 & 1.0 & 24.5 & 61.0 & 99.5 \\
    RACE              & 44.0 & 81.0 & 28.0 & 49.0 & 65.0 & 61.5 & 83.5 & 69.0 & 66.0 & 74.0 \\
    AutoDAN-R         & 96.5 & 95.5 & 88.5 & 100.0 & 98.0 & 96.0 & 89.5 & 94.0 & 94.5 & 96.0 \\
    PAIR              & 50.0 & 98.5 & 64.5 & 82.5 & 93.0 & 83.0 & 90.0 & 93.5 & 94.0 & 89.5 \\
    ReNeLLM           & 1.0 & 5.0 & 5.5 & 68.5 & 70.5 & 69.0 & 7.5 & 20.5 & 19.5 & 42.0 \\
    DrAttack          & 24.5 & 58.0 & 66.5 & 66.5 & 63.5 & 83.5 & 67.5 & 61.0 & 56.0 & 72.5 \\
    AutoDAN-Turbo     & 18.0 & 4.5 & 0.0 & 0.5 & 14.0 & 0.0 & 4.5 & 11.0 & 0.0 & 0.0 \\
    CipherChat        & 9.5 & 2.5 & 3.0 & 97.5 & 77.5 & 86.5 & 75.0 & 6.5 & 23.5 & 59.0 \\
    CodeAttack        & 41.5 & 92.5 & 44.5 & 83.5 & 83.5 & 79.0 & 73.5 & 86.5 & 89.5 & 87.0 \\
    Multilingual      & 3.5 & 0.5 & 3.0 & 62.5 & 11.5 & 27.5 & 0.0 & 1.0 & 33.5 & 7.0 \\
    Jailbroken        & 21.0 & 58.5 & 64.5 & 99.0 & 95.5 & 78.0 & 0.0 & 20.0 & 3.5 & 25.5 \\
    ICA               & 53.5 & 99.0 & 97.0 & 99.0 & 98.0 & 83.5 & 1.0 & 63.0 & 1.5 & 95.5 \\
    FlipAttack        & 90.5 & 17.5 & 97.5 & 99.0 & 91.5 & 91.5 & 31.0 & 53.5 & 12.5 & 97.0 \\
    Mousetrap         & 93.0 & 96.0 & 97.5 & 100.0 & 97.0 & 91.5 & 3.5 & 98.5 & 12.5 & 97.5 \\
    Pre-fill          & 6.0 & 1.0 & 0.5 & 99.5 & 96.0 & 50.5 & 1.5 & 4.0 & 3.5 & 36.0 \\
    DeepInception     & 2.0 & 29.0 & 44.0 & 99.0 & 99.5 & 97.0 & 0.0 & 22.0 & 1.5 & 97.0 \\
    Crescendo         & 12.0 & 49.0 & 21.5 & 56.0 & 59.0 & 57.5 & 50.5 & 94.5 & 47.5 & 46.5 \\
    RedQueen          & 0.5 & 3.0 & 1.5 & 24.0 & 47.0 & 36.5 & 3.0 & 24.0 & 2.5 & 2.0 \\
    CoA               & 10.0 & 7.0 & 1.0 & 9.5 & 9.0 & 8.5 & 53.5 & 31.0 & 11.5 & 37.5 \\
    ActorAttack       & 42.5 & 35.5 & 19.5 & 70.0 & 76.5 & 54.0 & 42.0 & 76.5 & 64.5 & 53.0 \\
    Rainbow Teaming   & 7.0 & 3.5 & 16.0 & 2.0 & 18.5 & 25.5 & 14.5 & 0.5 & 96.5 & 31.0 \\
    X-Teaming         & 94.0 & 98.5 & 80.5 & 94.0 & 99.0 & 89.5 & 93.0 & 98.5 & 97.0 & 95.0 \\
    EvoSynth          & 100.0 & 100.0 & 100.0 & 100.0 & 100.0 & 100.0 & 100.0 & 100.0 & 100.0 & 100.0 \\
    \bottomrule
    \end{tabular}}
    \caption{Attack Performance across Different LLMs on HarmfulBench}
    \label{tab:attack_llm}
\end{table}

\textbf{In summary}, these results reveal that adversarial robustness is highly \textbf{attack-dependent} and \textbf{model-dependent}: models withstand weak prompts yet fail under adaptive ones; simultaneously, attack strategies exhibit polarized outcomes, often proving lethal to specific targets while completely failing on others. By aggregating performance across methodologies (Table~\ref{tab:attack_methods}), we find that multi-agent and optimization-based methods are among the most potent. Notably, multi-agent strategies like EvoSynth and X-Teaming generalize best, with EvoSynth achieving near-universal success, demonstrating that multi-agent, search-driven synthesis effectively bypasses static safety constraints. Beyond multi-agent settings, black-box optimization and LLM-driven refinement (\emph{e.g.}, AutoDAN-R, PAIR, GPTFuzzer) also achieve strong and broadly consistent performance, reinforcing that adaptivity and iterative feedback are key drivers of attack effectiveness. Meanwhile, structured obfuscation and logic nesting (\emph{e.g.}, Mousetrap) can remain highly effective on many models, showing that attacks do not need to be multi-agent to be high-impact. In contrast, heuristic, template-heavy, or shallow linguistic/encoding manipulations tend to exhibit high variance and unstable ASR performance across models; while they may succeed on specific targets, they often fail entirely on others, suggesting that contemporary safety training and filtering increasingly mitigate static jailbreak patterns. Finally, weaker multi-turn heuristics (\emph{e.g.}, RedQueen) generally underperform, indicating that simple context manipulation alone is often insufficient against modern alignment. \textbf{Overall, these results highlight a shift in the attack landscape: adaptive, iterative, and multi-agent strategies dominate, whereas static, single-turn, and template-based approaches are increasingly brittle.}

\subsection{Multi-dimensional Attack Analysis}
While Attack Success Rate (ASR) serves as the primary indicator of vulnerability, a practical red-teaming framework requires a holistic assessment. 
As previously established, an ideal attack method should not only achieve \textbf{high ASR} but also demonstrate \textbf{high efficiency, high stealthiness, and high diversity}. In this section, we dissect these critical dimensions to evaluate the trade-offs inherent in different attack paradigms and identify which methods best approximate this ideal balance.

\begin{figure}[ht]
    \centering
    \includegraphics[width=0.95\linewidth]{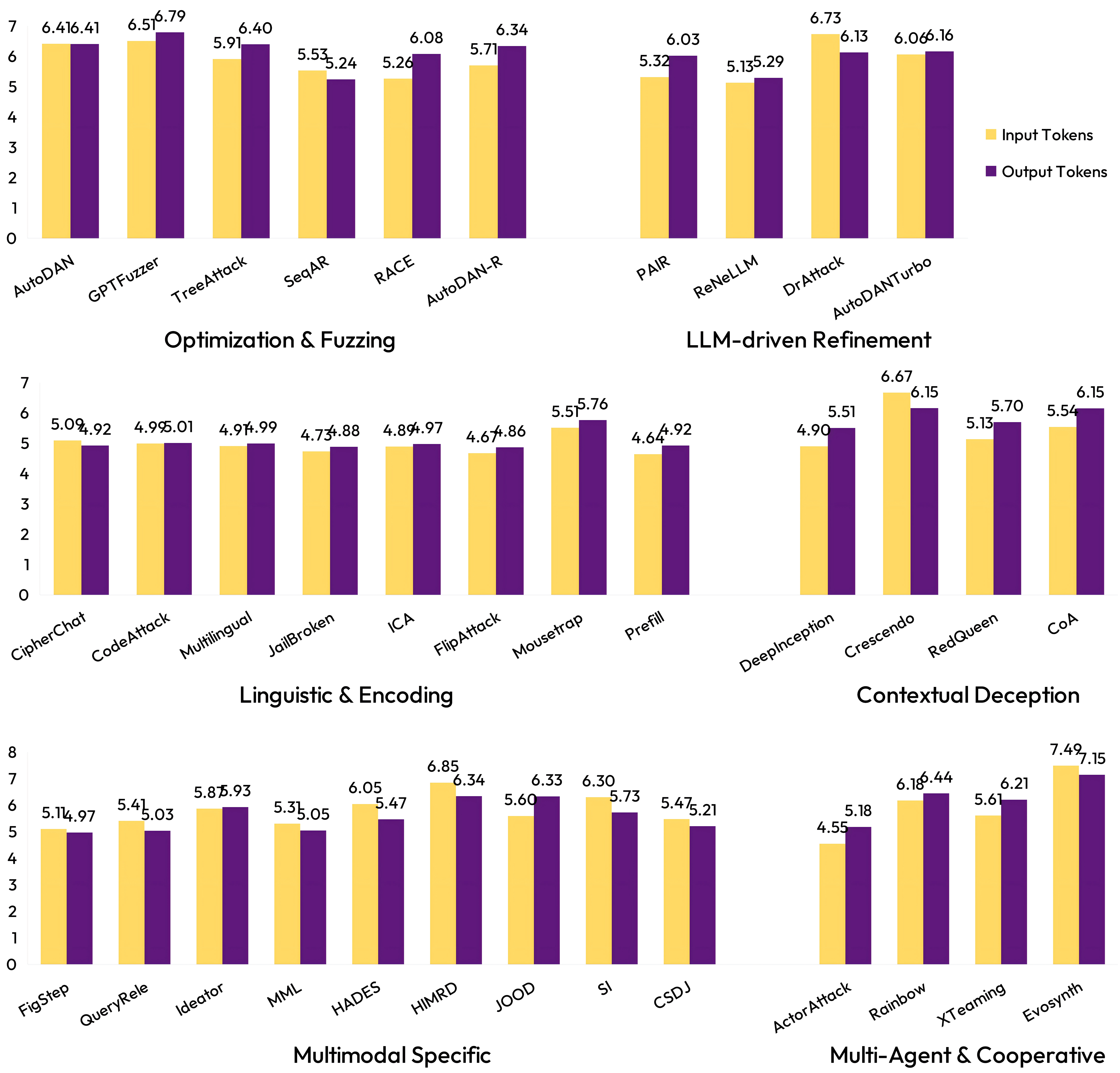}
    \caption{Computational cost of various attack methods against GPT-5.2, measured by input and output token usage (log$_{10}$ scale).}
    \label{fig:efficiency}
\end{figure}

\subsubsection{Attack Efficiency}
In addition to ASR, we evaluate the cost of running each attack using token and query statistics. For each method, we record input tokens and output tokens, total tokens (as a proxy for monetary/latency cost), and number of calls (proxy for rate-limit pressure and wall-clock time).

Figure~\ref{fig:efficiency} shows that GPTFuzzer is the most resource-intensive (9.44M total tokens; 9,705 calls), followed by DrAttack (6.75M total tokens; 3,224 calls), Crescendo (6.11M total tokens; 1,161 calls), and AutoDAN (5.14M total tokens; 9,406 calls). These approaches are dominated by iterative querying; GPTFuzzer is also notably output-heavy (6.19M output vs.\ 3.25M input tokens). Mid-cost methods such as TreeAttack, AutoDAN-Turbo, SI, X-Teaming, CoA, and IDEATOR reduce total tokens to roughly 1.6M--3.3M with fewer calls. In contrast, lightweight methods (\emph{e.g.}, FlipAttack, Prefill, JailBroken, ICA) enable low-budget sweeps with total tokens below 200k.

\begin{figure}[ht]
    \centering
    \includegraphics[width=0.85\linewidth]{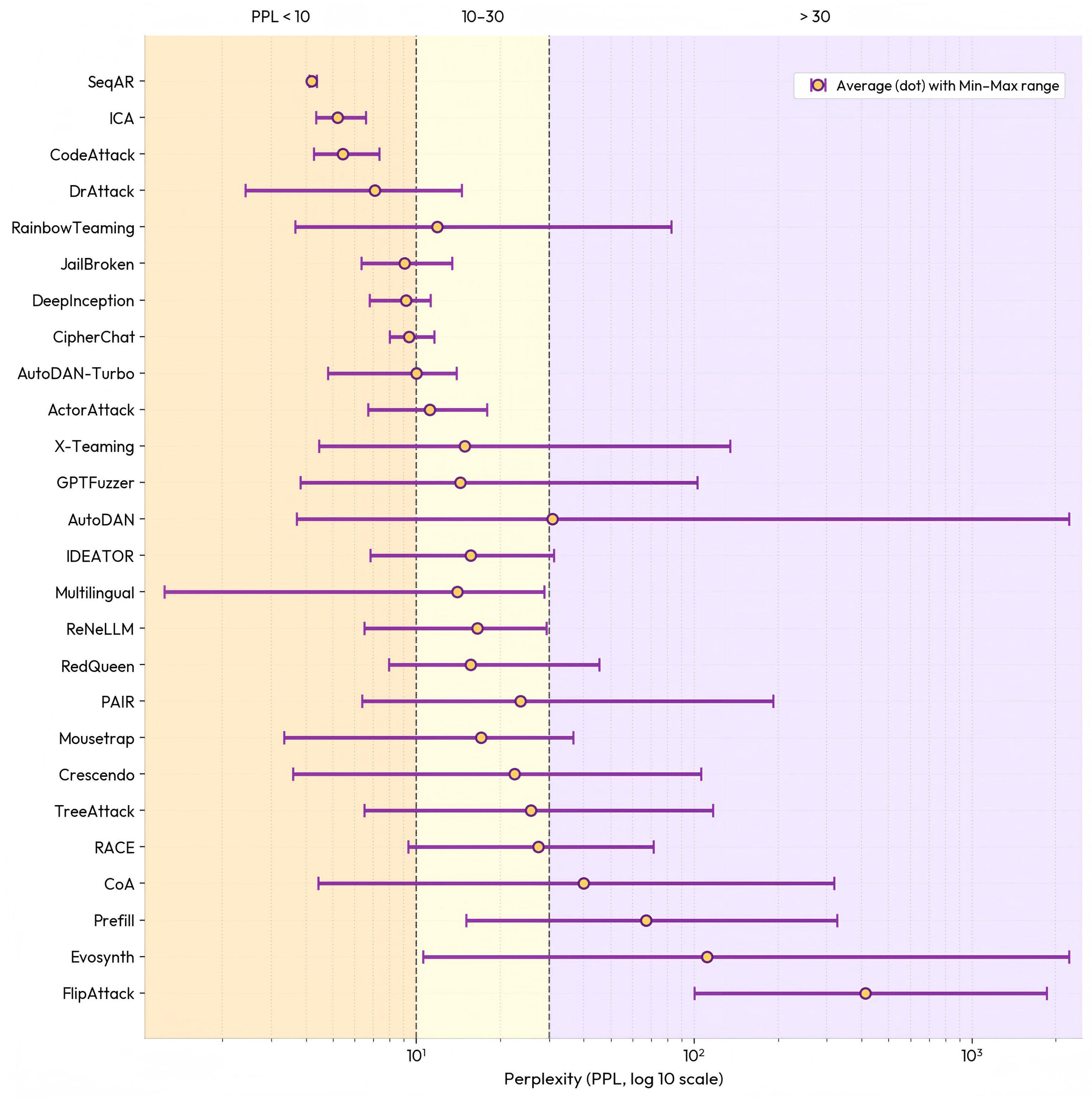}
    \caption{Attack stealthiness against GPT-5.2 measured by perplexity (PPL) on Qwen3-32B (log$_{10}$ scale). Lower PPL indicates more stealthy prompts. Shaded bands mark high ($<10$), moderate ($10$--$30$), and low ($>30$) stealthiness.}
    \label{fig:ppl}
\end{figure}

\subsubsection{Attack Stealthiness}
To evaluate the stealthiness of adversarial prompts in black-box attacks, we measure perplexity (PPL) with a Qwen3-32B base model as a proxy for linguistic naturalness, where lower PPL indicates prompts that are harder to detect by perplexity-based defenses. For white-box visual attacks, stealthiness is controlled via an $\epsilon$-bounded perturbation constraint. As shown in Figure~\ref{fig:ppl}, we observe three categories. \textbf{High-stealthiness attacks} (PPL $<$ 10) include SeqAR (4.18), ICA (5.18), CodeAttack (5.43), and DrAttack (7.09), which produce fluent prompts closely resembling natural language and are particularly concerning for evading detection. \textbf{Moderate-stealthiness attacks} (PPL 10--30) encompass iterative methods like RainbowTeaming (11.88), GPTFuzzer (14.35), X-Teaming (14.89), and multi-turn approaches like Crescendo (22.48) and PAIR (23.63), which balance exploration diversity with linguistic coherence. \textbf{Low-stealthiness attacks} (PPL $>$ 30) rely on obfuscation techniques, with FlipAttack (412.15) producing the least natural prompts due to character-level perturbations. Notably, attack effectiveness does not correlate with stealthiness: Mousetrap achieves 97.5\% ASR with moderate PPL (17.01), while SeqAR maintains excellent stealthiness (4.18) but only 25\% ASR, suggesting that defenders should employ multi-layered detection strategies combining semantic analysis with pplexity-based filtering.

\begin{figure}[ht]
    \centering
    \includegraphics[width=0.95\linewidth]{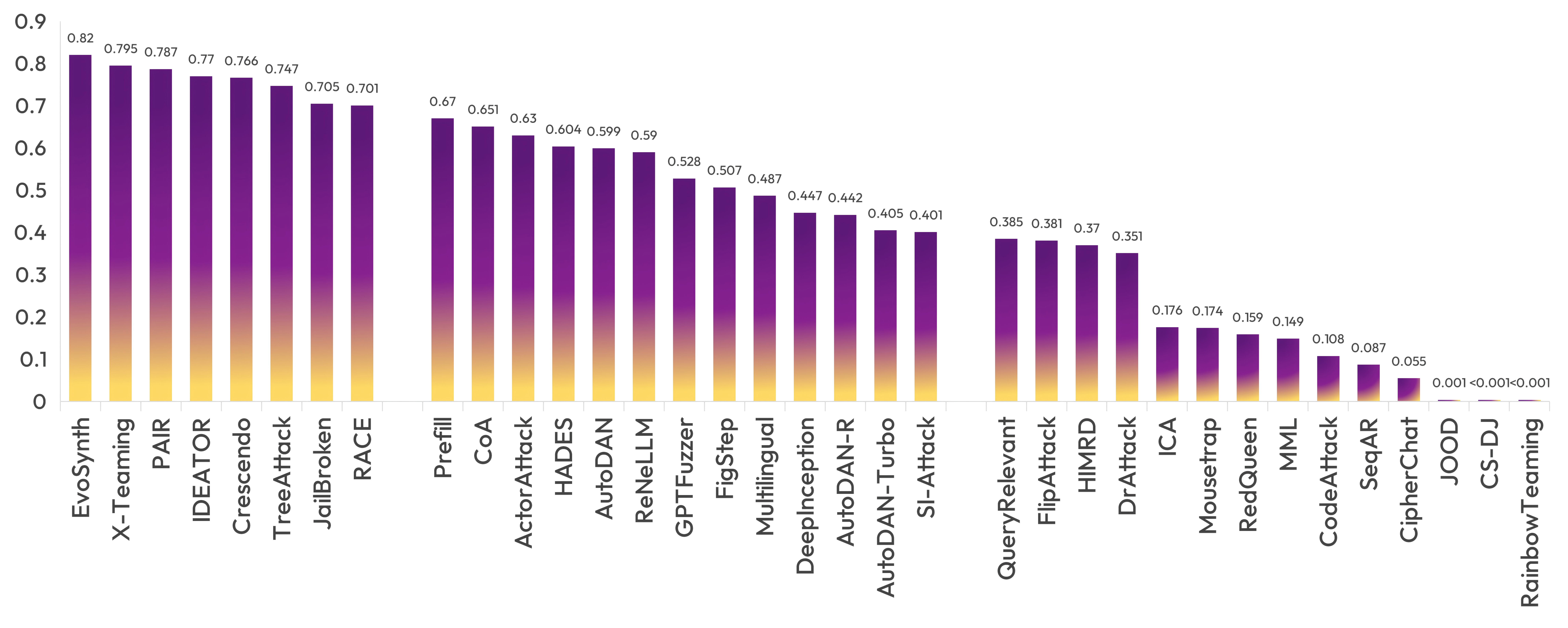}
    \caption{Diversity analysis of attack methods against GPT-5.2, quantified by the mean pairwise cosine distance between embeddings of successful adversarial prompts.}
    \label{fig:div}
\end{figure}

\subsubsection{Attack Diversity}
As shown in Figure~\ref{fig:div}, our analysis reveals three distinct categories. \textbf{High-diversity attacks} (Diversity $>$ 0.70) are dominated by multi-agent and iterative methods: EvoSynth (0.820), X-Teaming (0.795), PAIR (0.787), IDEATOR (0.770), and Crescendo (0.766) demonstrate the effectiveness of evolutionary and cooperative exploration mechanisms. \textbf{Moderate-diversity attacks} (0.40--0.70) include genetic algorithms (AutoDAN: 0.599, GPTFuzzer: 0.528) and contextual methods (CoA: 0.651, HADES: 0.604), where mutation operators introduce variation but often converge to similar patterns. \textbf{Low-diversity attacks} (Diversity $<$ 0.40) are characterized by template-based or encoding-constrained methods: CipherChat (0.055), CodeAttack (0.108), and SeqAR (0.087) produce structurally similar outputs due to their deterministic transformation schemes. Notably, RainbowTeaming, CS-DJ, and JOOD achieve near-zero diversity ($<$0.001), generating virtually identical prompts across different queries.

Interestingly, diversity does not always correlate with ASR. Mousetrap achieves 97.5\% ASR on GPT-5.2 despite low diversity (0.174), while high-diversity methods like TreeAttack (11.0\% ASR) explore broader but less effective attack strategies. This trade-off suggests that comprehensive red-teaming should combine high-diversity methods for vulnerability discovery with targeted low-diversity attacks for exploiting known weaknesses.

\section{Conclusion}\label{sec:conclusion}
In this work, we introduced \textbf{OpenRT}, a unified and extensible framework designed for comprehensive red-teaming evaluation of both MLLMs. By integrating 37 diverse attack methods, the framework provides a comprehensive tool for evaluating model safety, offering a standardized platform for benchmarking multiple models and attack strategies. Our large-scale evaluation of 20 advanced models revealed significant safety vulnerabilities in state-of-the-art systems, demonstrating that current safety mechanisms are often ineffective against a variety of adversarial techniques. OpenRT not only highlights persistent gaps in model defenses but also serves as a foundational infrastructure for future research in adversarial robustness. Looking ahead, we aim to expand OpenRT's capabilities by integrating emerging attack paradigms, enhancing support for additional modalities, and fostering community-driven evolution, ultimately helping to bridge the gap between perceived and actual safety in deployed AI systems.

\newpage

\newpage

\bibliographystyle{plain} 

\bibliography{main}

\newpage
\appendix

\section{Appendix: Usage and Extensibility}
This section explains how to use the \textbf{OpenRT} framework for security evaluations and how to extend it with your own attack methods.

\subsection{Basic Usage}

\paragraph{Installation}  
To install the framework, run:
\begin{figure}[ht]
\centering
\begin{minipage}{0.8\textwidth} 
\begin{minted}[fontsize=\small]{bash}
pip install -r requirements.txt
python setup.py install
\end{minted}
\captionof{listing}{Installing \textbf{OpenRT} from source.}
\label{lst:install}
\end{minipage}
\end{figure}

\paragraph{Running Experiments.}  
You can run experiments using configuration files or Python scripts.  
The configuration-driven method only requires a YAML file. For example:

\begin{figure}[ht]
\centering
\begin{minipage}{0.8\textwidth} 
\begin{minted}[fontsize=\small]{bash}
python main.py --config configs/autodan_turbo_experiment.yaml
\end{minted}
\captionof{listing}{Running an experiment from a YAML configuration file.}
\label{lst:run-config}
\end{minipage}
\end{figure}

Alternatively, you can set up the experiment programmatically:

\begin{figure}[ht]
\centering
\begin{minipage}{0.8\textwidth} 
\begin{minted}[fontsize=\small]{python}
from openrt import (
    OpenAIModel, StaticDataset, PAIR,
    LLMJudge, JudgeEvaluator, Orchestrator
)

# Initialize components
model = OpenAIModel(model_name="gpt-5.1", api_key="...")
helper_model = OpenAIModel(model_name="gpt-4o", api_key="...")
dataset = StaticDataset(prompts=["harmful query 1", ...])
judge = LLMJudge(judge_model=model, success_threshold=5)
attack = PAIR(model, helper_model, judge, max_iterations=5)
evaluator = JudgeEvaluator(judge=judge)

# Run experiment
orchestrator = Orchestrator(model, dataset, attack, evaluator)
metrics, results = orchestrator.run()

print(f"Attack Success Rate: {metrics.ASR:.2%}")
\end{minted}
\captionof{listing}{Programmatic setup of an OpenRT red teaming experiment.}
\label{lst:run-programmatic}
\end{minipage}
\end{figure}

\paragraph{Viewing Results}  
Results are saved in a folder with detailed logs:

\begin{figure}[ht]
\centering
\begin{minipage}{0.8\textwidth} 
\begin{minted}[fontsize=\small]{text}
results/
  baseline/
    gpt-5.1_20251207T074428Z/
      metrics/gpt-5.1_PAIR_metrics.json
      details/gpt-5.1_PAIR_results.jsonl
\end{minted}
\captionof{listing}{Example output directory structure and result files.}
\label{lst:results-structure}
\end{minipage}
\end{figure}

\subsection{Extending Attacks}

You can add new attack methods by implementing the \texttt{BaseAttack} interface and registering it with the framework. Here is an example of how to create a custom attack:

\begin{figure}[thp]
\centering
\begin{minipage}{0.8\textwidth} 
\begin{minted}[fontsize=\small]{python}
from openrt.attacks.base_attack import BaseAttack, AttackResult
from openrt.core.registry import attack_registry

@attack_registry.register("my_attack")
class MyAttack(BaseAttack):
    def __init__(self, model, max_iters=10, **kwargs):
        super().__init__(model, **kwargs)
        self.max_iters = max_iters
    
    def attack(self, target: str) -> AttackResult:
        for i in range(self.max_iters):
            adv_prompt = self._craft_prompt(target, i)
            response = self.model.query(adv_prompt)
            if self._is_successful(response):
                return AttackResult(
                    target=target,
                    success=True,
                    final_prompt=adv_prompt,
                    output_text=response,
                    method="my_attack",
                )
        return AttackResult(target=target, success=False)
\end{minted}
\captionof{listing}{Implementing and registering a custom attack in \textbf{OpenRT}.}
\label{lst:custom-attack}
\end{minipage}
\end{figure}

Once registered, your custom attack will be available for use in experiments: 
\begin{figure}[ht]
\centering
\begin{minipage}{0.8\textwidth} 
\begin{minted}[fontsize=\small]{yaml}
attack:
  name: "my_attack"
  args:
    max_iters: 15
\end{minted}
\captionof{listing}{Using the custom attack via configuration.}
\label{lst:custom-attack-config}
\end{minipage}
\end{figure}


    


\subsection{Configuration}
Complete experiments can also be declared purely through YAML configuration files, upon which the Orchestrator dynamically instantiates and wires together the corresponding components (model, dataset, attack, judge, and evaluator). An example configuration for a complete experiment is shown in Listing~\ref{lst:config}.

\begin{figure}[thp]
\centering
\begin{minipage}{0.8\textwidth} 
\begin{minted}[fontsize=\small]{yaml}
experiment_name: "Comprehensive_Safety_Evaluation"

model:
  name: "openai"
  args:
    model_name: "gpt-5.1"
    temperature: 0.7
    api_key: "${OPENAI_API_KEY}"

dataset:
  name: "jsonl"
  args:
    file_path: "data/advbench.jsonl"

attack:
  name: "autodan_turbo"
  args:
    epochs: 5
    warm_up_iterations: 2
    lifelong_iterations: 3
    break_score: 8.5

evaluator:
  name: "judge"
  args:
    judge:
      name: "llm_judge"
      args:
        success_threshold: 5
\end{minted}
\captionof{listing}{YAML configuration for a complete OpenRT red-teaming experiment.}
\label{lst:config}
\end{minipage}
\end{figure}

\end{document}